# SURVEY REPORT – STATE OF THE ART IN DIGITAL STEGANOGRAPHY FOCUSING ASCII TEXT DOCUMENTS

Khan Farhan Rafat
Department of Computer Science
International Islamic University
Islamabad, Pakistan

Muhammad Sher
Department of Computer Science
International Islamic University
Islamabad, Pakistan

*Abstract*— **Digitization of analogue signals has opened up new avenues for information hiding and the recent advancements in the telecommunication field has taken up this desire even further. From copper wire to fiber optics, technology has evolved and so are ways of covert channel communication. By "Covert" we mean "anything not meant for the purpose for which it is being used". Investigation and detection of existence of such cover channel communication has always remained a serious concern of information security professionals which has now been evolved into a motivating source of an adversary to communicate secretly in "open" without being allegedly caught or noticed.**

**This paper presents a survey report on steganographic techniques which have been evolved over the years to hide the existence of secret information inside some cover (Text) object. The introduction of the subject is followed by the discussion which is narrowed down to the area where digital ASCII Text documents are being used as cover. Finally, the conclusion sums up the proceedings.**

## I. INTRODUCTION

**Cryptography** derived from Greek, (where historian *Plutarch* elaborated on the use of **scytale** – *an encryption technique via transposition*, a thin wooden cylinder, by a general for writing message after wrapping it with paper, to decrypt the message, one needs to wrap that piece of paper again on the scytale to decrypt the message [41].), focuses on making the secret information unintelligible.

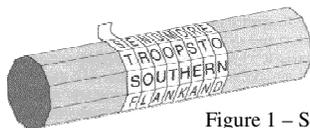

Figure 1 – Scytale [44]

Information Hiding Men's quest to hide information is best put in words [2] as "we can scarcely imagine a time when there did not exist a necessity, or at least a desire, of transmitting information from one individual to another in such a manner as to elude general comprehension".

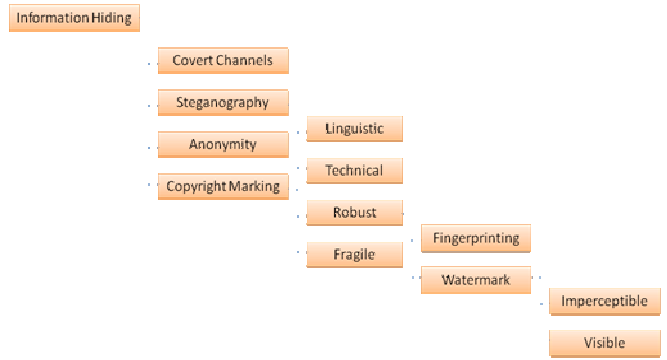

Figure 2 – Classification of Information Hiding based on [1]

While discussing information hiding, we mainly come across people from two schools of thought. One votes for making the secret information unintelligible (encryption) [5] whereas the other like *E*neas the Tactician, and John Wilkins [4][5] are in favor of hiding the existence of the information being exchanged (steganography) because of the fact that the exchange of encrypted data between Government agencies, parties etc. has its obvious security implications.

- Covert/Subliminal Channel A communication channel which is not explicitly designed for the purpose for which it is being used [6][7] e.g. using TCP & IP header for hiding and sending secret bits etc.
- Steganography is derived from the Greek words , 'steganos' and 'graphie', $\sigma\tau\epsilon\gamma\alpha\nu\acute{o}\varsigma \ \gamma\rho\alpha\varphi\epsilon\iota\nu$ [8] which means Covered Writing/Drawing.

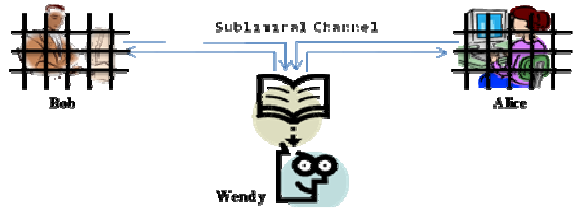

Figure 3 – Prisoner's Problem

The *classic* model for invisible communication was first proposed by Simmons [3][4] as the ***prisoners' problem*** who argued by assuming, for better understanding, that Alice and Bob, who have committed a crime, are kept in separate cells of a prison but are allowed to communicate with each other via a warden named Wendy with the restriction that they will not encrypt their messages and that the warden can put them in isolated confinement on account of any suspicious act while in communication. In order to plan an escape, they now need a subliminal channel so as to avoid Wendy's intervention.

Following is an example from [34] where in World War I, German Embassy in Washington (DC) sent the following telegram messages to its Berlin headquarters (David Kahn 1996):

"**P**RESIDENT'S **E**MBARGO **R**ULING **S**HOULD **H**AVE **I**MMEDIATE **N**OTICE. **G**RAVE **S**ITUATION **A**FFECTING **I**NTERNATIONAL **L**AW. **S**TATEMENT **F**ORESHADOWS **R**UIN **O**F **M**ANY **N**EUTRALS. **Y**ELLOW **J**OURNALS **U**NIFYING **N**ATIONAL **E**XCITEMENT **I**MMENSELY.

**A**PPARENTLY **N**EUTRAL'S **P**ROTEST **I**S **T**HOROUGHLY **D**ISCOUNTED **A**ND **I**GNORED. **I**SMAN **H**ARD **H**IT. **B**LOCKADE **I**SSUE **A**FFECTS **P**RETEXT **F**OR **E**MBARGO **O**N **B**YPRODUCTS, **E**JECTING **S**UETS **A**ND **V**EGETABLE **O**ILS." [34]

By concatenating the first character of every word in the first message and the second character of every word in the second message the following concealed message is retrieved:

**"PERSHING SAILS FROM NY JUNE I"** [34]

| Advantages | Disadvantages |
|---|---|
| Does not require a device for computational purposes. | Does not follow Kerchoff's principle. Requires voluminous data for trespassing and Needs careful generation and crafting of cover text for hiding bits. |

*A.* **Terminology:** By convention, the object being used to hide information within it is called cover-text. A variety of media such as text, image, audio etc. depicted in [9][10][11][42] are used to hide secret information within its body. After embedding of secret information, the resultant object is referred to as stegotext/stego-object. According to [12] the algorithms by virtue of which secret information is embedded in the cover-text at the sending end, and gets extracted out of stego-text at the receiving end constitutes a stego-system. The secret key involved in information exchange [13] via private and public key Steganography is referred to as stego-key.

*B.* **Model:** Though different in their approach, Steganography and Cryptography go well together when it comes to information security. The evolution of digital technology (which is a continuous process) has brought significant change in the methodologies being used / preferred earlier for hiding information. As now we opt for a blend of these two techniques added with compression, to attain a near to perfect security solution having '*no-compromise on security*' as our slogan. Mathematical modeling of Figure-4 follows:

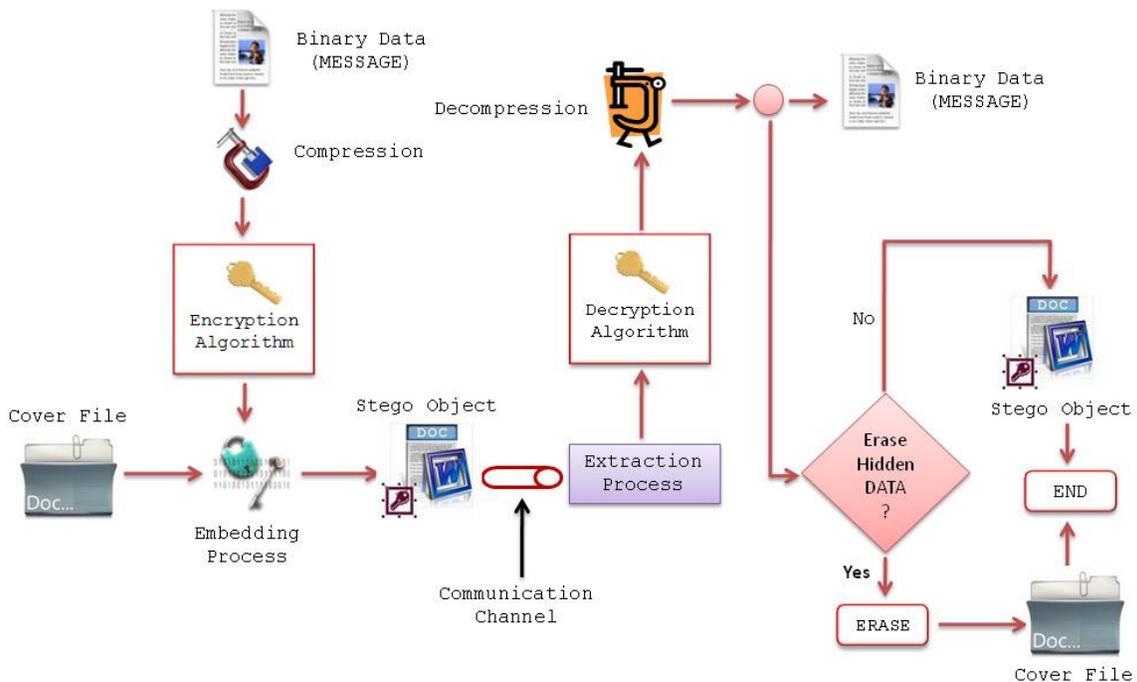

Figure 4 – Preferred Stegosystem

At present Internet spam is (and can be) a potential candidate to be used as cover for hiding information.

- *Encoding Process:*
  $$Ὸ = η\ (o, ġ, Ҡ)$$

Where**:**

**η** is the function which operates on the Cover **'o'** to embed compressed and encrypted data 'ġ' using the Stego-Key $K$ to produce the Stego-Object **'Ό'**.

ġ = ē (ć, ǩ) 'Encrypting the compressed message (ć) with secret key.

ē is the encryption function that takes the compressed data ć for encryption using symmetric key ǩ.

ć = c(M) 'Compressing the secret message (M) using appropriate algorithm.

- **Decoding Process:**
$$o = Ę ('Ό', ġ, K)$$
Where**:**

Ę is the function which operates on the stegocover object **'Ό'** to decompress and decrypt the data indicated by function 'ġ' using the Stego-Key $K$ and extract the secret information.

ġ = đ (ć, ǩ) 'Decrypting the decompressed data (ć) with secret key ǩ.

đ is the decryption function that takes the compressed data ć for decryption using symmetric key ǩ.

ć = c (M') 'Decompressing the hidden message (M') using appropriate algorithm.

C. **Categorization:** Steganography is broadly categorized in [2][7] as:

- **Linguistic** Variety of techniques (such as discussed in [15][16][17][18][19]), takes advantage of the syntax and semantics of Natural Language (NL) for hiding information. However, the earliest form probably of which is acrostic. Giovanni Boccaccio's *Amorosa visione* is considered as the world's hugest acrostic [20, pp. 105–106] (1313–1375).
- **Technical** This technique is broader in scope which is not confined to written words, sentences or paragraphs alone but involves some kind of tool, device or methodology [16] for embedding hidden information inside a cover, particularly in its regions / areas which remain unaffected by any form of compression.

D. **Categorization of Steganographic Systems based on techniques** as explained in [8] is as under:

- **Substitution systems** Redundant parts of cover get replaced with secret information.
- **Transform domain techniques** Transform space of the signal is used for embedding information such as frequency domain.
- **Spread Spectrum technique** uses the conventional approach as is done Telecommunication sector where a signal is spread over a range of frequencies.
- **Statistical methods** encode information by changing several statistical properties of a cover and use hypothesis testing in the extraction process.
- **Distortion techniques** store information by signal distortion and measure the deviation from the original cover in the decoding step.
- **Cover generation methods** encode information in the way a cover for secret communication is created.

E. **Types of Embedding Applications**

Another important pre-requisite for covert channel communication is the availability of some type of application embodying some algorithm/technique for embedding secret information inside the cover. The birth of Inter and Intranet has given way to a multitude of such applications where information hiding finds its vitality as was never before.

Following is a brief review as of [36] of such applications which are differentiated according to their objectives:

- **Non-Repudiation, Integrity and Signature Verification**: Cryptography concerns itself with making the secret information un-intelligible by using the techniques of confusion and diffusion as suggested by Shannon, to ensure integrity of the message contents. Public key cryptography is a preferred way of authenticating the sender of the message (i.e. the sender/signature is genuine / non-repudiation). This, however, becomes challenging when the information is put on line as a slight error in transmission can render the conventional authentication process as a failure; hence now there are applications for automatic video surveillance and authentication of drivers' licenses etc.
- **Content Identification**: By adding content specific attributes such as how many times a video is being watched or a song is played on air; one can judge the public opinion about it.
- **Copyright Protection**: The most debated, popular and yet controversial application of information hiding is copyright protection as it is very easy to have an exact replica of a digital document / item and the owner / holder of the document can own or disown its rights. One such popular incident occurred in 1980's when British Prime Minister being fade up about the leakages of important cabinet documents got the word processor modified to automatically encode and hide the user's information within word spacing of the document to pin-point the culprits. In the early 1990's, people begin to think about digital watermarking for copyright compliance.
- **Annotating Database**: It is not un-common for large audio / video databases to have text or speech etc.

captions which can easily be got embedded inside the relevant database to resist against various signal processing anomalies.

- **Device control**: Human audio / video perceptions are frequently being exploited by the vendors in designing their information hiding techniques. In one such reported technique a typical control signal, embedded in a radio signal broadcasted by a FM radio station was used to trigger the receiver's decoder.
- **In-Band captioning**: Just as it is not un-common to embed data in audio-visual streams; so can be the case where data for various services launched by Telecom Operators can be embedded in television and radio signals.
- **Traitor Tracing**: Here distinct digital signatures are embedded and the number of copies to be distributed is limited. The unauthorized usage of the document can then be traced back by the intended recipient.
- **Media Forensics**: The tempered media gets analyzed by experts to identify the tempering and the portions which have been affected by it but not throw light as to how the tempering is done.

F. **Types of Steganographic Systems:** According to [8] Steganographic Systems can be segregated as:
- **Pure Steganography (PS):** Weakest, as is based on the assumption that parties other than the intended ones are not aware of such type of exchange of secret information.
- **Secret Key Steganography (SKS):** In this technique, both the sender and receiver share or have agreed on a common set of stego-keys prior to commencement of secret communication. The secret information is embedded inside the cover using the pre-agreed stego-key and gets extracted out at the receiving end by reversing the embedding process. The advantage lies in the fact that an adversary needs to apply brute force etc. attack to get the secret information out of the cover which require resources such as computational power, time, and determination.
- **Public Key Steganography (PKS)** As the name indicates, public key steganography use a pair of Public and Private Keys to hide secret information. The advantage of this technique is that an attacker first needs to come up with a public and private key-pair and then the decoding scheme to extract the hidden information out of the cover. The key benefit of this technique is its robustness in execution and easy of key management.

G. **Models for Steg-Analysis**

- **Blind Detection Model**
  This model is the counterpart of cryptanalysis and analyzes the stego-object without any prior knowledge about the technology and the type of the media (cover) being used in the concealing process.

- **Analytical Model**
  The stego-object is analyzed in terms of its associated attributes such as stego-object type, format etc. [21] and thereafter on the basis of the data gathered relevant known steg-analysis tools are used to extract hidden bits and derive the meaning out of it.

## II. Related Work

This section covers a literature review of the recently published text-based steganographic techniques such as use of acronyms, synonyms; semantics to hide secret bits in English Text in paras *A – E*, format specific techniques are discussed in paras *F – K* while ways of hiding secret bits in TCP and IP header are elaborated in para *L* respectively:

A. **Acronym**

According to the definition at [43] "*an acronym (pronounced AK-ruh-nihm, from Greek acro- in the sense of extreme or tip and onyma or name) is an abbreviation of several words in such a way that the abbreviation itself forms a pronounceable word. The word may already exist or it can be a new word. Webster's cites SNAFU and radar, two terms of World War Two vintage, as examples of acronyms that were created*".

Mohammad Sirali-Shahreza and M. Hassan Shirali-Shahreza have suggested the substitution of words with their abbreviations or *viza viz* in [40] to hide bits of secret message. The proposed method works as under:

Table 1

| Acronym | Translation |
|---------|-------------|
| 2l8 | Too late |
| ASAP | As Soon As Possible |
| C | See |
| CM | Call Me |
| F2F | Face to face |

*If a matched word/abbreviation is found then the bit to be hidden is checked to see if it is under column '1' or '0' and based on its value (i.e., 0 or 1), word/abbreviation from the corresponding column label is substituted in the cover message, in case of otherwise the word/abbreviation is left unchanged. The process is repeated till end of message.*

| Advantages |
|---|
| Speedy |
| Flexibility: More words / abbreviation pairs can be added. |
| Technique can be applied in variety of fields such as science, medicine etc. |

| Disadvantage |
|---|
| The main drawback lies in the static word/abbreviation substitution where anyone who knows the algorithm can easily extract the hidden bits of information and decode the message which is against Kerckhoff's Principle which states that the security of the System should lie in its key, where the algorithm is known to public. |

*B.* **Change of Spelling**

Mohammad Shirali-Shahreza in his paper [23] proposed a method of exploiting the way; words are spelled differently in British and American English, to hide bits of secret information. The procedure for concealment explained below, is the same as that of para *A*, where the words spelled in British and American English, are arranged in separate columns as shown in Table 2.

Table 2

| American Spelling | British Spelling |
|---|---|
| Favorite | Favourite |
| Criticize | Criticise |
| Fulfill | Fulfil |
| Center | Centre |

*The column containing British Spelling is assigned label '1' while that containing American Spelling is assigned label '0'. The information to be hidden is converted into bits. The message is iterated to find differently spelled words matching to those available in pre-defined list (Table 2 refers).*

*If a matched word is found then the bit to be hidden is checked to see if it is under column '1' or '0' and based on its value (i.e., 0 or 1), word spelled in American or British English from the corresponding column label is substituted in the cover message. Words not found in the lists are left unchanged. The process is repeated till end of the message.*

| Advantage | Disadvantages |
|---|---|
| Speedy | Language Specific |
|  | Non-adherence to Kerckhoff's principle. |

*C.* **Semantic Method**

Mohammad Sirali-Shahreza and M. Hassan Shirali-Shahreza in [24] have used those English Language words whose synonym exists.

The authors had arranged words (having synonym) in one column while corresponding synonyms were placed in another column (Table 3 refers) and followed the procedure explained below:

*The column containing words/Translation is assigned label '1' while that containing acronyms is assigned label '0' and the information to be hidden is converted into bits. The message is iterated to find words/abbreviations matching to those available in pre-defined list (Table 1 refers).*

Table 3

| Word | Synonym |
|---|---|
| Big | Large |
| Small | Little |
| Chilly | Cool |
| Smart | Clever |
| Spaced | Stretched |

*If a matched word / synonym is found then the bit to be hidden is checked to see if it is under column '1' or '0' and based on its value (i.e., 0 or 1), word or synonym from the corresponding column label is substituted in the cover message. Words not found in the lists are left unchanged. The process is repeated till end of the message.*

| Advantage | Disadvantages |
|---|---|
| Speedy | Language Specific |
|  | Non-adherence to Kerckhoff's principle. |
|  | Only one synonym is taken in the substitution table. |

*D.* **Miscellaneous techniques**

The authors in [31] have given a number of idiosyncrasies ways that are / can be used for hiding secret message bits, such as by introducing modification or injecting intentional grammatical word/sentence errors to the text. Some of the suggested techniques / procedures which can be employed in this context include:

- *Typographical errors - "tehre" rather than "there".*
- *Using abbreviations / acronyms - "yr" for "your" / "TC" in place of "Take Care".*
- *Transliterations – "gr8" rather than "great".*
- *Free form formatting - redundant carriage returns or irregular separation of text into paragraphs, or by adjusting line sizes.*
- *Use of emoticons for annotating text with feelings - ":)" to annotate a pun.*
- *Colloquial words or phrases - "how are you and family" as "how r u n family".*
- *Use of Mixed language - "We always commit the same mistakes again, and 'je ne regrette rien'!".*

| Advantages | Disadvantages |
|---|---|
| More variations for hiding information. | Eye catching. |
| More computations required. | Can draw suspicion. |

*E.* **Enhanced Steganography in SMS**

In his paper at [35] the author has suggested an enhancement in an existing steganographic system [22] by taking care of the limitations of the existing techniques discussed in paras *A – D* which work as under:

*In this enhanced technique, words and their corresponding abbreviations are grouped under two columns. The column containing words is labeled as '1' and that containing abbreviations is labeled as '0 (Table 4 refers)'. Depending on the input 128-bit stego-key bits and the value of the first stego-key byte, words and their corresponding abbreviations are swapped so that the two columns now contain a mix of words and abbreviations.*

Table 4

|   | 1 | 0 |
|---|---|---|
| 0 | Too late | 2l8 |
| 1 | ASAP | As Soon As Possible |
| 0 | See | C |
| 1 | CM | Call Me |
| 1 | F2F | Face to face |

- **Bits Embedding Process**_*A 128-bit Linear Feedback Shift register (LFSR), initialized using the same stego-key, serves as a pseudo random bit generator, the first 128 bits of which are discarded before use. The output bits from the LFSR get XoR-ed with the bits of the message. Based on the resultant bits of the XoR operation, words or abbreviations corresponding to column labels replaces the contents of the original message.*

The embedding and extraction processes are depicted diagrammatically in Figure 5 and 6 respectively:

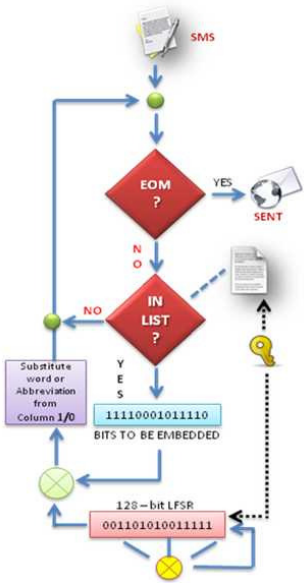 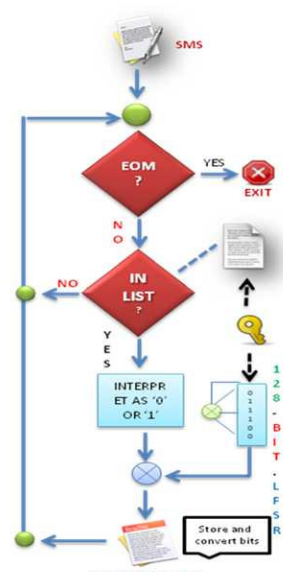

Figure 5.Embedding Process      Figure 6.Extraction process

- **Bits Extraction Process**_*It is just the reverse of the Bits embedding process, where after Initialization; the hidden bits are first extracted and then get XOR-ed with the output of 128-bit LFSR. The resultant bits are concatenated and passed through a transformation which translates the string of bits into their equivalent ASCII character i.e. secret message text.*

| Advantages |
|---|
| Adherence to Kerchoff's Principle |
| Shanon's principles of confusion & diffusion |
| Secret bits are encrypted before being embedded in the cover makes the system secure, as the enemy will have to perform additional efforts of decrypting the bits without the knowledge of key. |

| Advantages |
|---|
| The 128-bit LFSR used for encryption with a non repeated key has rendered the system as OTP. |
| The algorithm can be extended to the desktop, PDA platforms. |
| The algorithm is language independent. |
| Adding compression before encryption can hide more secret bits in the cover. |

| Disadvantage |
|---|
| Slightly slower (in fractions) than its predecessor technique. |

*F.* **MS Word Document**

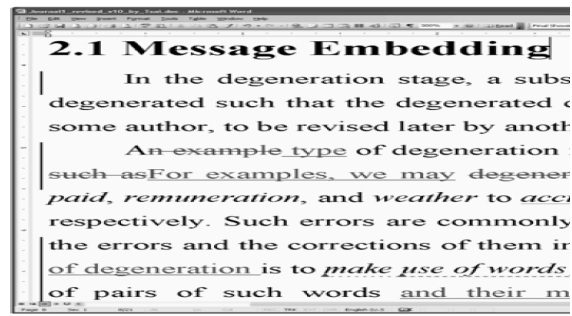

Figure 7 – MS Word for Steganography

The author at [32] has made use of change tracking technique of MS Word for hiding information, where the stego-object appeared to be a work of collaborated writing. As shown in Figure 7, *the bits to be hidden are first embedded in the degenerated segments of the cover document which is followed by the revision of degenerated text thereby imitating it as being an edited piece of work.*

| Advantage | Disadvantage |
|---|---|
| Easy to use as most users are familiar with MS word. | Easily detectable as MS Word has built in spell check and Artificial Intelligence (AI) features. |

*G.* **HTML Tags**

The author of publication at [38] elaborates that *software programs like '**Steganos** for **Windows**' uses gaps i.e. space and horizontal tab at the end of each line, to represent binary bits ('1' and '0') of a secret message. This, however, adds visibility when the cover document is viewed in MS Word with visible formatting or any other Hex-Editor* e.g.:

<html>( )->->( )->
<head>( )->( )( )->
[Text]()()->
</head>->()->->
</html>()->()->

Where **( )** represents Space and '**->**' denotes Horizontal Tab.

The above example indicates hiding of secret bits '100101001…' as per analogy explained above.

*Spaces are also inserted in between TAGS to represent a hidden bit '0' or '1'. The above example indicates hiding of secret bits '1001001010' as per analogy explained.*

*Later in the discussion, the author proposed the use of line shift; interpreted (in hex) as 0xA0, 0xD0 in Windows and as 0xA0 in Unix Operating System to translate these as '1' and '0'. A majority of text editors can interpret the two codes for line shift without ambiguity; hence it is a comparatively secure way to hide secret information.*

The author of [39] has shown ways where *HTML TAGS can also be manipulated to represent hidden bit '0' or '1'.*

| Advantage | Disadvantages |
|---|---|
| Works well for HTML documents as regards on screen visibility. | Visibility/Eye catching in case of TEXT documents. |
| | Increase in Stego-cover File size. |
| | Non-adherence to Kerchoff's principle. |

### H. XML Document

![Figure 8: Data Hiding in XML document]

Figure 8: Data Hiding in XML document

XML is a preferred way of data manipulation between web-based applications hence techniques have been evolved as published in [26] for hiding secret information within an XML document. The user defined tags are used to hide actual message or the placement of tags represents the corresponding secret information bits. *One such technique places hidden text bytes sequentially in Tags as shown in Figure 8.*

| Advantage | Disadvantages |
|---|---|
| XML is widely acceptable tool for information exchange which makes the task of its Steg-analysis difficult. | Eye catching. |
| | Increased Stego-cover File size. |
| | Non adherence to Kerchoff's principle. |

### I. White Spaces

W. Bender, D. Gruhl, N. Morimoto, and A. Lu in [25] have discussed a number of steganographic techniques for hiding data in a cover, where one of the methods *places one or two spaces after every terminated sentence of the cover file/text to represent a secret bit '0' or '1' as the case may be.*

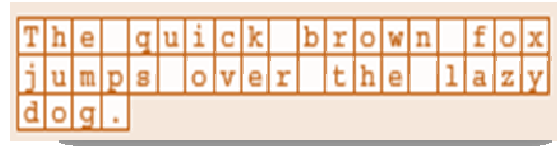

Figure 9:   Original Text [25]

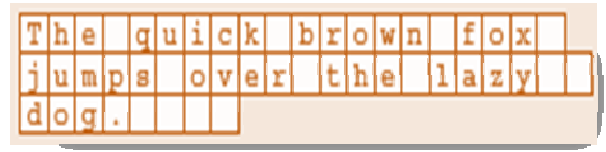

Figure 10:   Stego-Text [25]

*Another discussed technique includes hiding data by text justification as shown in* Figure 11.

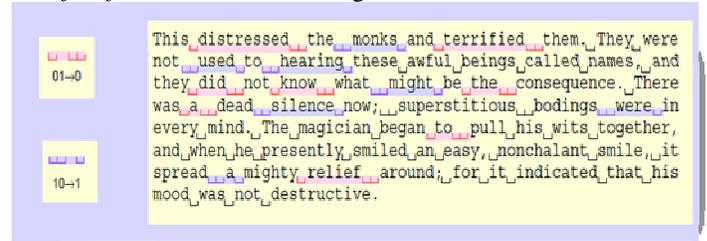

Figure 11 -Text from 'A Connecticut Yankee' in King Arthur's Court by Mark Twain [25]

| Advantage | Disadvantages |
|---|---|
| Normally passes by undetected. | Violates Kerckhoff's Principle. |
| | Increases cover text size. |

### J. Line Shifting

Printed text documents can also be manipulated as an image and subjected to steganographic techniques such as discussed in [28][29] by *slight up/down lifting of letters from baseline or right/left shifting of words within a specified image/page width*, etc.

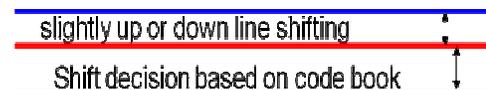

Figure 12 [28]

*This increase/decrease in line height or the increase decrease in space between words by left / right shifting can be interpreted as binary bits '0' or '1' accordingly to hide secret information.*

Figure 13 [28]

| Advantage | Disadvantages |
|---|---|
| Difficult to detect in the absence of original text. | Looses format if the document is saved as text. |

### K.   Feature Coding

The steganographic method in [30] *hides the secret information bits by associating certain attributes to the text characters like changing font's type, its size, color, by underlining it or using strike-through etc.*

e.g.   Steganography is the art of hiding secret information.

| Advantages | Disadvantages |
|---|---|
| More variations for hiding information. | Eye catching. |
| More computations required. | Can draw suspicion. |

### L.   IPv4 and Transport Layer

Richard Popa [33] has analyzed a variety of steganographic techniques and among the techniques discussed, those related to Internet protocol (IP) and Transmission control protocol (TCP) are discussed here. *Figure 14 shows how the IP (version 4) header is organized. Three unused bits have been marked (shaded) as places to hide secret information. One is before the DF and MF bits and another unused portion of this header is inside the* Type of service *field which contains two unused bits (the least significant bits).*

Figure 14 [33]

*Every TCP segment begins with a fixed-format 20-byte header. The 13th and 14th bytes of which are shown in Figure 15. The 6-bit field not used, indicated in shade, can be used to hide secret information.*

Figure 15 [33]

| Advantage | Disadvantage |
|---|---|
| Due to enormous packet flow almost unlimited amount of secret bits can be exchanged via these techniques. | Loss of packets may render undesirable results. |

## III.   Conclusion

This paper presents a survey on a data hiding technique called 'Steganography', the terminology, the model, its types, and two types of attacks on any Steganographic system. This is followed by a discussion on various text-based data-hiding techniques where the primary focus remained on recently proposed/developed Steganographic techniques.

**Secure e-Governance** An essential feature of e-government includes secure transmission of confidential information via computer networks where the sensitivity of some information may fall equivalent to a level as that of national security. Every e-government has its own network but cannot ignore the Internet which by far, is the cheapest means of communication for common people to interact with the Government. The data on Internet, however, is subjected to hostile attacks from Hackers etc. and is therefore a serious e-government concern. In his paper at [37] the author has emphasized on the importance of steganography for use in e-Government and discussed that *Governments, seek and had sought consultation and help from cryptographers and have invested huge amounts of time and funds in getting developed specially designed information security systems to strengthen data security. In today's' world, cryptography alone is just not an adequate security solution. With the increase in computation speed, the old techniques of cryptanalysis are falling short of expectations and will soon be out-dated. Steganology – that encompasses digital data hiding and a detection technique has gained considerable attention now days. It appears to be a powerful opponent to cryptology and offers promising technique for ensuring seamless e-security.*

From the discussion, it is apparent that ensuring one's privacy has remained and will always remain a serious concern of Security frontiers. The innocent carrier i.e., text document (ASCII text format), will continue to retain its dominance in time to come for being the preferred choice as cover media, because of zero overhead of metadata with its body.

[42] **Chapman, Mark**. *A Software System for Concealing Ciphertext as Innocuous Text*, Hiding the Hidden: http://www.NICETEXT.com/NICETEXT/doc/ thesis.pdf.1997

[43] http://searchcio-midmarket.techtarget.com/ sDefinition/0,,sid183_gci211518,00.html

[44] http://mail.colonial.net/~abeckwith/images/ scytale.gif

## AUTHORS PROFILE

**KHAN FARHAN RAFAT** has completed his Ph.D. course work under supervision of Professor Dr. Muhammad Sher, International Islamic University, Islamabad.

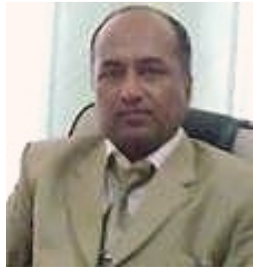

He has twenty years R & D experience in the field of Information and communication Security ranging from formulation and implementation of security policies, evolution of new and enhancement of existing security related Projects to software development etc.

**Professor Dr. MUHAMMAD SHER** is Head of Department of Computer Science, International Islamic University, Islamabad, Pakistan. He did his Ph.D. Computer Science from TU Berlin, Germany, and specialized in Next Generation Networks and Security.

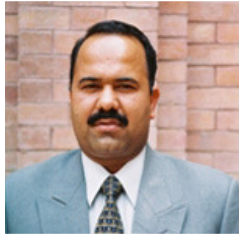

He has vast research and teaching experience and has a number of international research publications to his credit.